\documentclass{article}
\usepackage{caption}
\usepackage{subcaption}
\usepackage{authblk}
\usepackage{amsmath}
\usepackage{graphicx}
\usepackage{amssymb}
\usepackage{booktabs}
\usepackage{amsmath}
\usepackage{bm}
\usepackage{siunitx} 

\begin{document}

\title{A Neural-Operator Surrogate for Platelet Deformation Across Capillary Numbers}

\author[1]{Marco Laudato\footnote{Corresponding author; laudato@kth.se}}

\affil[1]{FLOW Research Center, Department of Engineering Mechanics, KTH Royal Institute of Technology, Stockholm, SE-10044, Sweden.}

\maketitle

\begin{abstract}
Reliable multiscale models of thrombosis require platelet-scale fidelity at organ-scale cost, a gap that scientific machine learning has the potential to narrow. We train a DeepONet surrogate on platelet dynamics generated with LAMMPS for platelets spanning ten elastic moduli and capillary numbers (0.07 - 0.77). The network takes in input the wall shear stress, bond stiffness, time, and initial particle coordinates and returns the full three-dimensional deformation of the membrane. Mean-squared-error minimization with Adam and adaptive learning-rate decay yields a median displacement error below 1\%, a 90th percentile below 3\%, and a worst case below 4\% over the entire calibrated range while accelerating computation by four to five orders of magnitude. Leave-extremes-out retraining shows graceful extrapolation: the held-out stiffest and most compliant platelets retain sub-3\% median error and an 8\% maximum. Error peaks coincide with transient membrane self-contact, suggesting improvements via graph neural trunks and physics-informed torque regularization. These results classify the surrogate as high-fidelity and position it for seamless coupling with continuum CFD, enabling platelet-resolved hemodynamic simulations in patient-specific geometries and opening new avenues for predictive thrombosis modeling.

\end{abstract}

\section{Introduction}
Scientific machine learning is challenging the long-standing computational barrier that separated molecular-scale platelet physics from organ-scale blood-flow models.
Physics-informed~\cite{karniadakis2021physics} and operator-learning architectures (e.g., DeepONet~\cite{lu2021learning}) can learn the nonlinear map between membrane-level mechanics and flow conditions directly from high-fidelity particle- or molecular-dynamics data and then inject that knowledge into a continuum computational fluid dynamics (CFD) solver at virtually zero computational cost.
Recent platelet-specific implementations~\cite{laudato2024high,laudato2025neural} show that a single DeepONet evaluation reproduces full platelet dynamics at molecular level with sub-percent error while accelerating the micro-solver by four to five orders of magnitude, potentially enabling simulations with clinically realistic platelet counts and geometries that were previously infeasible~\cite{macraild2024accelerated}.

The application of such surrogates in a multi-scale model loop can be impactful for thrombosis modeling~\cite{xu2008multiscale}, where clot initiation depends on nanometer-scale activation~\cite{tomaiuolo2017regulation} yet is modulated by vessel-scale haemodynamics~\cite{zhang2002platelet}.
Bridging these two complementary descriptions of thrombosis is crucial~\cite{zhang2016multi}. Continuum-mechanics models excel at capturing haemodynamics in both idealized~\cite{laudato2023buckling, laudato2024analysis, laudato2024sound} and complex, patient-specific geometries~\cite{gutierrez2024decoding,karmonik2008computational,sundstrom2023machine,bornemann2024relation}, making them indispensable for clinically relevant scenarios, while molecular-dynamics simulations deliver high-fidelity in the mechanochemical behaviour of individual blood components, including receptor binding, membrane deformation, and chemical signalling~\cite{zhang2021predictive, wang2023multiscale,gupta2021multiscale}.

By coupling operator-learned surrogates like DeepONet into a multiscale loop, it is possible to unify the geometric flexibility of continuum approaches with the atomistic accuracy of molecular models, thereby obtaining a holistic (and computationally tractable) framework for thrombus initiation and progression.
The fidelity of such multiscale framework is bounded by the surrogate’s ability to emulate the underlying molecular dynamics; rigorous benchmarking across the governing dimensionless groups is therefore mandatory.
In this study we focus on the capillary number, $Ca^*$, that can be interpreted as the ratio of viscous to elastic forces~\cite{kruger2016effect} acting on the platelet.
This number governs platelet and red blood cell deformation and margination in shear flows, as summarized in Table~\ref{tab:capillary-number-examples} and extensively characterized in prior work on blood-cell suspensions.

\begin{table}[ht]
\centering
\begin{tabular}{>{\raggedright\arraybackslash}p{0.5\textwidth}|c|c}
\hline
Application & $Ca^*$ [-] & Reference\\
\hline
Red-blood-cell organization in straight micro-capillaries & $0.07$ & \cite{liao2022flow}\\
\hline
Platelet mechanotransduction  & 0.2 (derived) & \cite{abidin2023microfluidic}\\
\hline
Red-blood-cell diffusivity and margination & 0.3 & \cite{malipeddi2021shear}\\
\hline
Platelet - Red-blood-cell interaction & 0.7 & \cite{vahidkhah2013hydrodynamic}\\
\hline
\end{tabular}
\caption{Examples of dimensionless capillary numbers in micro- and nano-scale blood-flow studies, related to the range considered in this work.}
\label{tab:capillary-number-examples}
\end{table}

As $Ca$ increases, platelets undergo tank-treading motions and flattening that enhance their lateral drift toward the vessel wall (margination), whereas at low $Ca$ they retain a more rigid, discoidal shape with limited near-wall transport and adhesive surface area~\cite{dynar2024platelet}.
These deformation-driven changes also modulate platelet–red cell collision rates and shear-gradient diffusivity, both of which sensitively scale with $Ca^*$ and determine how frequently platelets encounter and adhere to the endothelium~\cite{tuna2024platelet}.
In thrombosis, where clot initiation is based on the interplay between shear-driven transport and receptor–ligand binding kinetics, capturing the full spectrum of $Ca^*$-dependent behaviors is essential.
Systematic studies of $Ca^*$ therefore provide the mechanistic underpinning for training and validating neural-operator surrogates like DeepONet across the physiologically relevant deformation regimes encountered in vivo.

In this study, we quantify the accuracy of a DeepONet-based surrogate in capturing the time‐dependent deformation of a platelet suspended in Couette flow. The platelet is represented by a network of roughly 20 000 particles linked by harmonic springs, with variations in the capillary number achieved by adjusting the spring constant. At each time step, the neural operator receives the current flow conditions and membrane parameters and returns the platelet’s instantaneous particle configuration. To assess the surrogate's robustness, we systematically evaluate model performance across a range of capillary numbers, conducting both interpolation tests within the training regime and extrapolation tests beyond it. This framework allows us to determine the surrogate’s predictive limits and to establish guidelines for its deployment in multiscale thrombosis simulations.

The platelet’s particle-based model and the governing haemodynamics are presented in Sec.~\ref{sec:lammps}. In Sec.~\ref{sec:deeponet}, we detail the neural operator architecture applied in this work. Finally, Sec.~\ref{sec:capillary} discusses the surrogate’s error across varying capillary numbers.

\section{Particle–based platelet model}
\label{sec:lammps}
The learning dataset is generated with \textsc{LAMMPS}~\cite{thompson2022lammps} by coupling dissipative particle dynamics (DPD)~\cite{groot1997dissipative} for the surrounding blood flow to a spring-network representation of the platelet, following the workflow established in our previous studies.
In our initial work~\cite{laudato2024high}, we sampled 101 discrete shear stresses (50–250 Pa) but recorded only the end‐state platelet shapes after one Jeffery orbit, whereas a subsequent study~\cite{laudato2025neural} resolved the full time evolution at a single shear stress (50 Pa).
In the present work, we capture complete deformation trajectories across 10 different capillary numbers, obtained by varying the harmonic‐bond stiffness.
In total, 101 distinct time instants are sampled for the 10 values of the harmonic-bond elastic constant $K$ (Table~\ref{tab:capillary_sweep}). The resulting dimensional capillary number

\begin{equation}
    Ca^{*} = \frac{\mu\,\dot{\gamma}\,a}{G_s}
    \label{eq:Ca}
\end{equation}

\noindent
covers the physiologically relevant range \(0.07 < Ca^{*} < 0.7\), where \(\mu\) is the blood viscosity, \(\dot{\gamma}\) the imposed shear rate, \(a\) the platelet’s characteristic radius, and \(G_s\) the in-plane surface shear modulus linked to \(K\).

\begin{table}[ht]
    \centering
    \begin{tabular}{@{}c c c@{}}
        \toprule
        Shear stress $\sigma$ [Pa] & Bond constant $K$ [N\,m$^{-1}$] & Resulting $Ca^{*}$ \\ 
        \midrule
        50
            & $K = \{0.0003-0.0030\}$ 
            & 0.07–0.70 \\ 
        \bottomrule
    \end{tabular}
    \caption{Parameter sweep defining the capillary-number dataset.
    The bond constant $K$ is varied from 0.0003 to 0.0030 N\,m$^{-1}$ in steps of 0.0003 N\,m$^{-1}$.}
    \label{tab:capillary_sweep}
\end{table}

\subsection{Simulation domain and boundary conditions}
\label{ssec:domain}
\begin{figure}[h!]
    \centering
    \includegraphics[width=0.75\linewidth]{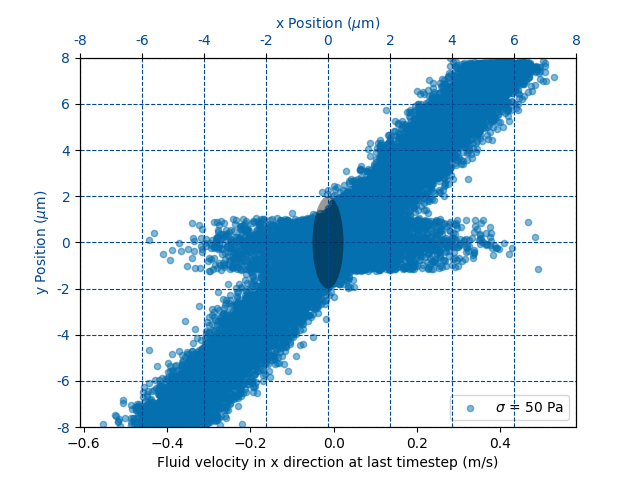}
    \caption{Steady-state distribution of DPD fluid particle velocities around the platelet in its initial position (dark ellipsoid at center, for reference). Each blue dot represents the x-component of a fluid particle’s velocity after relaxation, plotted against its y-coordinate, under a Couette flow of $\sigma = 50\,$Pa. The clear gap around the ellipsoid arises from the no-penetration boundary condition.}
    \label{fig:simulation_domain}
\end{figure}

The computational box measures \(16\times16\times8~\si{\micro\meter}\) (Fig.~\ref{fig:simulation_domain}). Opposing translational velocities \(U\) are prescribed at the top and bottom walls to realize a Couette flow; the required velocity for a target shear stress \(\sigma\) follows the linear relation

\begin{equation}
    U \;=\; \frac{\sigma\,L}{2\mu},
\end{equation}

\noindent
with \(L = 16~\si{\micro\meter}\) the wall spacing.
Periodic boundary conditions are enforced in the remaining directions. 
The time step \(\Delta t = 2.4\times10^{-10}\,\si{\second}\) ensures numerical stability across all \(Ca^{*}\) cases, and each run is advanced for one Jeffery period, providing a complete flip-and-deform cycle of the platelet.

\subsection{Platelet membrane model}
\label{ssec:membrane}

\begin{figure}[h!]
    \centering
    \includegraphics[width=0.31\linewidth]{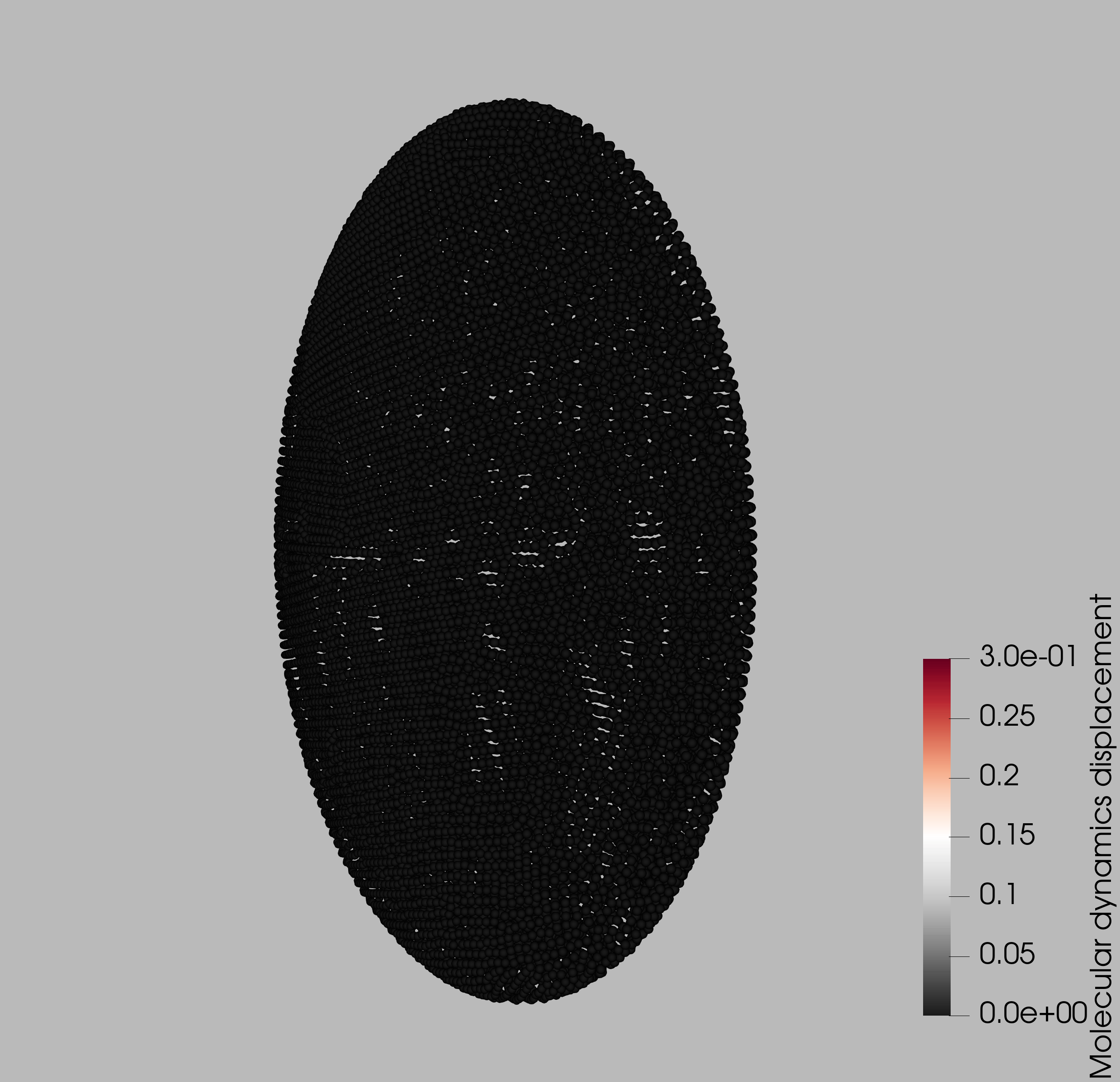}
    \includegraphics[width=0.31\linewidth]{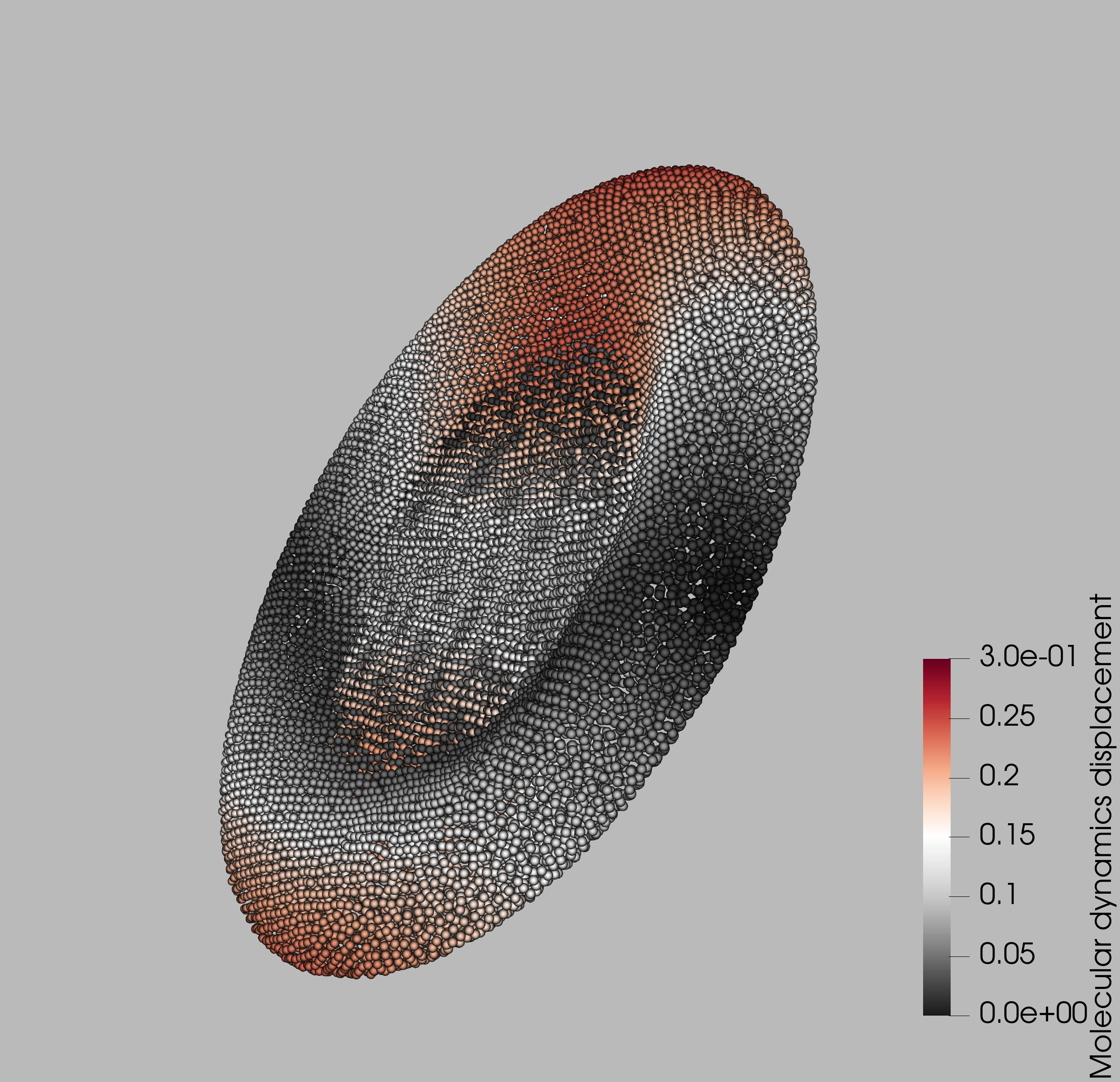}
    \includegraphics[width=0.31\linewidth]{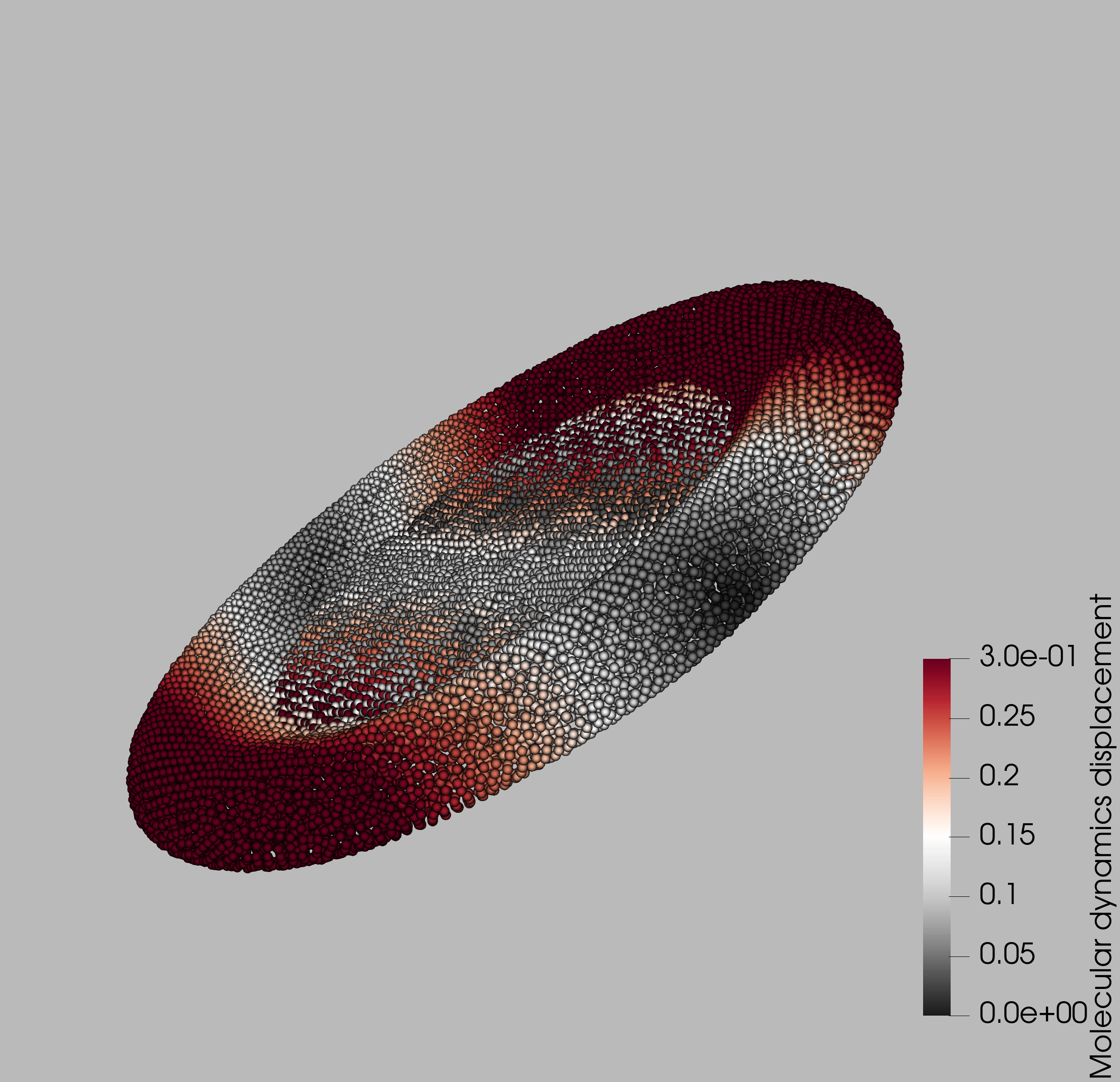}
    \caption{Time‐resolved deformation of a platelet under Couette flow from LAMMPS simulations. Left, center, and right panels correspond to early, intermediate, and late time points within one Jeffery orbit. The color scale denotes the magnitude of each particle’s displacement vector $\mathbf{u}(t)$ normalized by the maximum initial displacement, i.e.\ $\|\mathbf{u}(t)\|/\max\limits_{p}\|\mathbf{u}_{p}(0)\|$.}    \label{fig:lammps_platelet}
\end{figure}

The platelet is initialized as a hollow ellipsoid \(4\times4\times1~\si{\micro\meter}\) discretised into \(\approx 18\,000\) particles by 3-D Delaunay triangulation.  Nearest neighbors are linked by harmonic bonds,
\begin{equation}
    U_{\text{harm}} = \sum_{\text{bonds}} K\,(r-r_0)^2,
\end{equation}
where \(K\) is varied to realize the desired \(Ca^{*}\) range.
Non-bonded blood–platelet interactions use a truncated Lennard-Jones potential
plus dissipative and random DPD forces to enforce a no-slip, non-penetrating interface~\cite{zhang2014multiscale}.
The resulting dynamics is depicted in Fig.~\ref{fig:lammps_platelet}.

For every value of $K$, we record the particle positions at each time step, yielding \(\mathcal{O}(10^{6})\) labeled states per trajectory. The resulting database underpins the DeepONet-based surrogate, which is trained on 80\% subset of the trajectories and validated both \emph{interpolatively} (within the spanned parameter grid) and \emph{extrapolatively}, as detailed in Sec.~\ref{sec:deeponet}.

\section{Neural Operator Surrogate Model}
\label{sec:deeponet}
The surrogate model is trained to emulate the operator
\begin{equation}
    \mathcal{G}:\; \bigl(\mathbf{x}_0,\;t;\;\sigma,\;K\bigr)
    \;\longmapsto\;
    \mathbf{x}(t),
\end{equation}

\noindent
where $\mathbf{x}_0=(x_0,y_0,z_0)$ denotes the initial position of a membrane particle, $t$ is the simulation time, $\sigma=50$ Pa is the imposed wall shear stress, and $K$ the harmonic–bond elastic constant that sets the capillary number. 
For every membrane particle in the platelet, the network returns the particle’s instantaneous coordinates $\mathbf{x}(t) = (x(t),y(t),z(t))$.
Thus the surrogate learns the full, time–resolved deformation field of a platelet given its initial geometry and the governing flow–material parameters.

\begin{figure}[h!]
    \centering
    \includegraphics[width=0.45\linewidth]{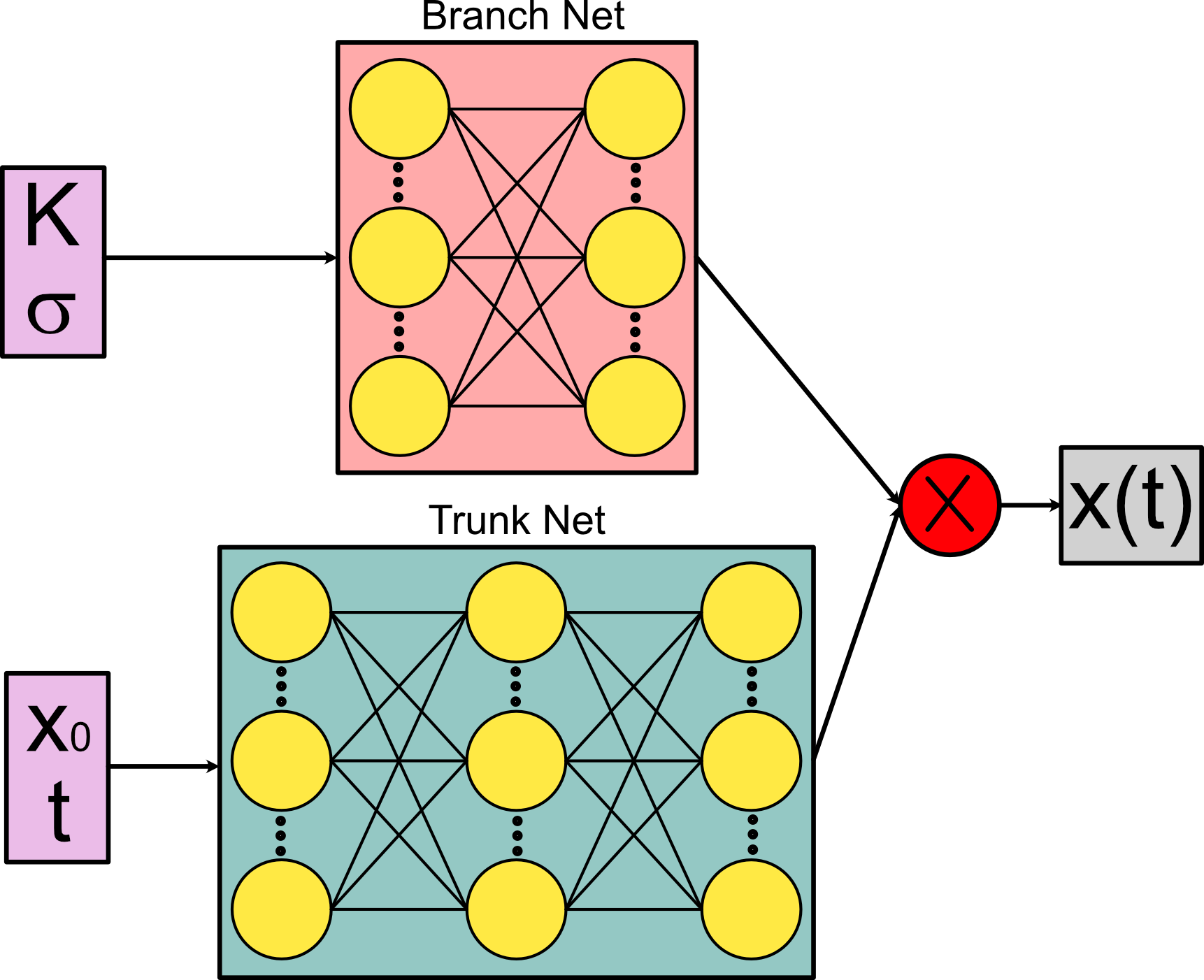}    
    \includegraphics[width=0.51\linewidth]{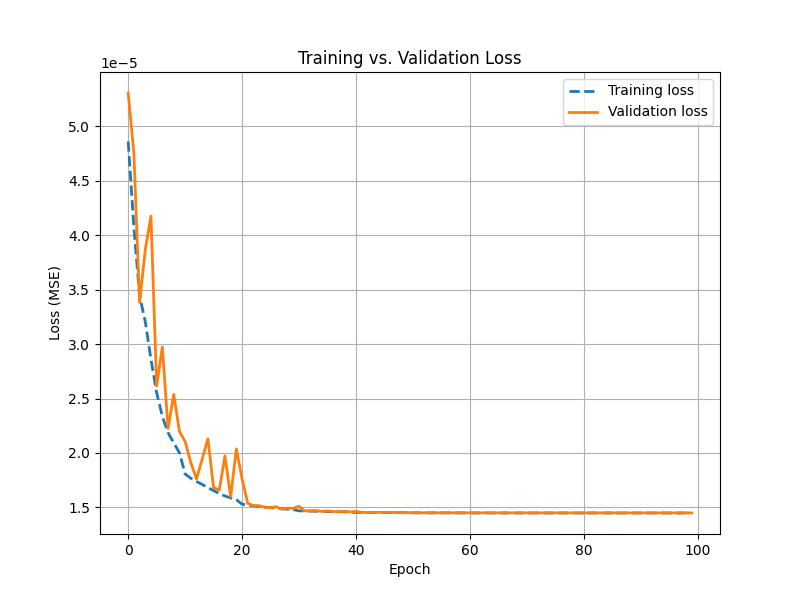}
    \caption{\textit{(left panel)} Schematics of the DeepONet architecture; \textit{(right panel)} Convergence of the DeepONet: training and validation mean-squared error (MSE) drop rapidly within the first $\sim25$ epochs, with no divergence between the two curves, indicating good generalization.}
    \label{fig:lossVsVal}
\end{figure}

The neural network implementation (see Fig.~\ref{fig:lossVsVal}, left panel) follows the architecture of DeepONet~\cite{lu2021learning}.
Both the \textit{branch net} and the \textit{trunk net} are fully-connected neural networks.
The branch net contains 2 hidden layers with 32 and 16 nodes, while the trunk net contains 3 hidden layers with 32, 32, and 16 nodes.
The latent space dimension (which is the same for both networks) is given by 32 nodes.
The ReLU activation function is employed at the nodes of the hidden layers, whereas the linear activation function is reserved for the outputs.
The scalars $(\sigma,K)$ forms the branch input, while $(\mathbf{x}_0,t)$ is supplied to the trunk.
The output of the two network is processed via a inner product and the corresponding  output is the particles absolute position \(\mathbf{x}(t)\).

The training dataset is drawn from the LAMMPS simulations (Sec.~\ref{sec:lammps}) and split 90\% for training and 10\% for validation. 
The network weights are optimized using the Adam algorithm over 100 epochs, with an initial learning rate of $10^{-3}$ that is adaptively reduced to $10^{-7}$ on plateau.
The loss function is the mean squared error between the network prediction and ground truth, and we track the mean absolute error on the validation set to detect overfitting.
The resulting convergence is shown in Fig.~\ref{fig:lossVsVal} (right panel).

\begin{figure}[h!]
    \centering
    \includegraphics[width=0.49\linewidth]{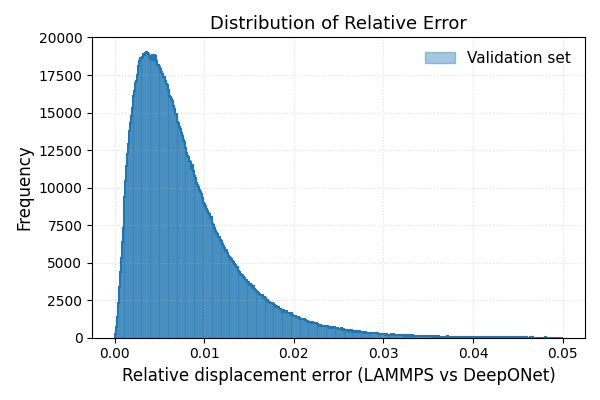}
    \includegraphics[width=0.49\linewidth]{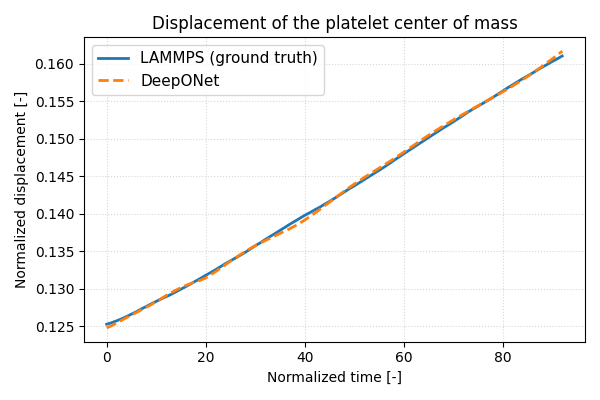}
    \caption{\textit{(left panel)} Histogram of relative displacement errors on the validation set, showing that mode of DeepONet predictions incur less than 1\% error relative to LAMMPS. \textit{(right panel)} Time series of the platelet’s normalized center-of-mass displacement demonstrating near-perfect agreement between DeepONet and LAMMPS over the full time evolution.}
    \label{fig:histo_CoM}
\end{figure}

A preliminary performance assessment compares the relative displacement error of DeepONet predictions (computed on the held-out validation set) with the corresponding LAMMPS ground truth.
As shown in Fig.~\ref{fig:histo_CoM} (left), the error distribution is sharply peaked below 1\% and exhibits a long, thin tail extending to 5\%.
The right panel of Fig.\ref{fig:histo_CoM} demonstrates that the surrogate accurately tracks the platelet’s center-of-mass displacement throughout the Couette‐flow trajectory.
Achieving sub-percent error while accelerating the time‐dependent deformation by several orders of magnitude highlights the surrogate’s promise for multiscale blood‐flow modeling.

\section{Per-capillary Number Analysis}
\label{sec:capillary}
The aim of this section is to quantify how the surrogate error varies with membrane stiffness, expressed through the capillary number ${Ca^{\ast}}(\sigma,K)$. 
A pragmatic training strategy was adopted: rather than excluding an entire stiffness level, the network was trained on a uniformly random \(90\,\%\) subset of \(\bigl\{\mathbf{x}_0,t,\sigma,K\bigr\}\) tuples, leaving the remaining \(10\,\%\) for validation.  Consequently, every value of \(K\) appears in the optimizer mini-batches, but only a sparse selection of time–particle samples at that \(K\) influences the weights.
The per-\(K\) study presented below evaluates the fully trained network on complete platelet trajectories, \emph{including time instants and  particle indices never seen during training}, thus providing a strict interpolation benchmark across the full \(Ca^{\ast}\) span.

While a leave-one-\(K\) cross-validation would yield a pure extrapolation metric, it would also entail retraining ten separate
models on data sets exceeding \(10^8\) samples each, which is beyond the scope of the present work.
The interpolation results reported here therefore represent the surrogate’s expected accuracy when embedded in multiscale  hemodynamic simulations that operate within the calibrated capillary-number range.

The per–capillary-number results are summarized in Table~\ref{tab:per-k-error}, which reports the median, 90th‐percentile, and maximum relative errors between DeepONet predictions and LAMMPS simulations.
Relative errors increase with capillary number (as lower bond stiffness produces larger deformations) but even in the most extreme cases the maximum error stays below 4\%, demonstrating the surrogate’s high fidelity.
Interestingly, the stiffest case (lowest $Ca^*$) shows a slightly higher maximum error than some intermediate stiffnesses. 
This is attributable to its location at the edge of the training domain, where the model effectively operates in a quasi‐extrapolative regime.
A focused assessment of extrapolation performance is presented in Sec.~\ref{ssec:extrapolation}.

\begin{table}[h!]
    \centering
    \begin{tabular}{c|c|c|c|c}
         \hline
         $K$ [N/m]& $Ca^*$ [-]& Rel. error (50\%)& Rel. error (90\%) & Rel. error (max)\\
         \hline
         0.0003 &0.7698 &0.0094 &0.0224 &0.0389 \\
         0.0006 &0.3849 &0.0076 &0.0179 &0.0314 \\
         0.0009 &0.2566 &0.0070 &0.0165 &0.0288 \\
         0.0012 &0.1925 &0.0066 &0.0158 &0.0270 \\
         0.0015 &0.1540 &0.0062 &0.0152 &0.0262 \\
         0.0018 &0.1283 &0.0058 &0.0145 &0.0244 \\
         0.0021 &0.1100 &0.0057 &0.0131 &0.0212 \\
         0.0024 &0.0962 &0.0056 &0.0122 &0.0208 \\
         0.0027 &0.0855 &0.0058 &0.0122 &0.0214 \\
         0.0030 &0.0770 &0.0056 &0.0126 &0.0228 \\
        \hline
    \end{tabular}
    \caption{Validation-set relative displacement errors for each bond stiffness \(K\) and corresponding capillary number \(\mathrm{Ca}^*\). Shown are the median (50th percentile), 90th percentile, and maximum errors of DeepONet predictions versus LAMMPS ground truth.}
    \label{tab:per-k-error}
\end{table}

\begin{figure}[h!]
    \centering
    \includegraphics[width=0.32\linewidth]{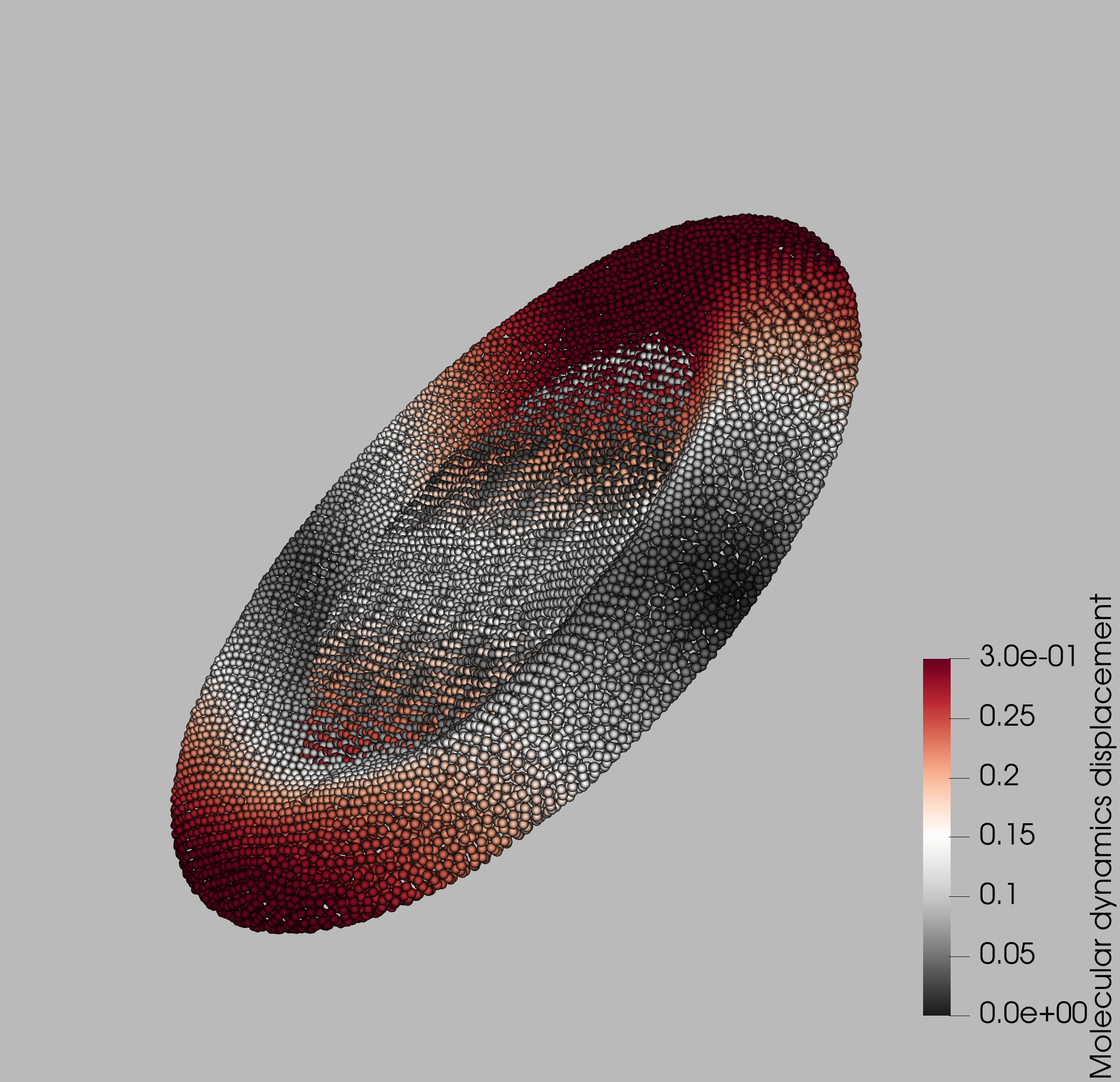}
    \includegraphics[width=0.32\linewidth]{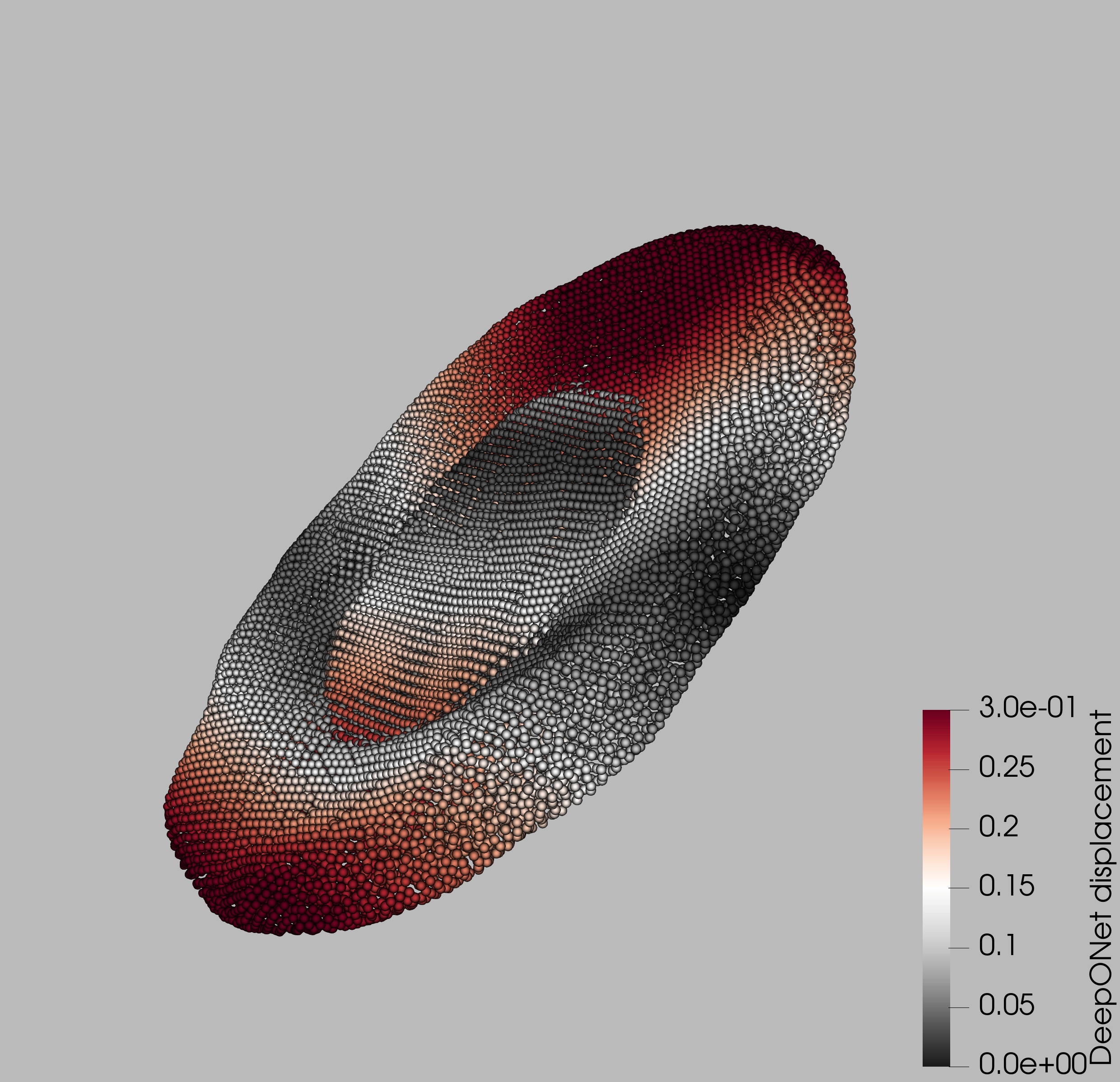}
    \includegraphics[width=0.32\linewidth]{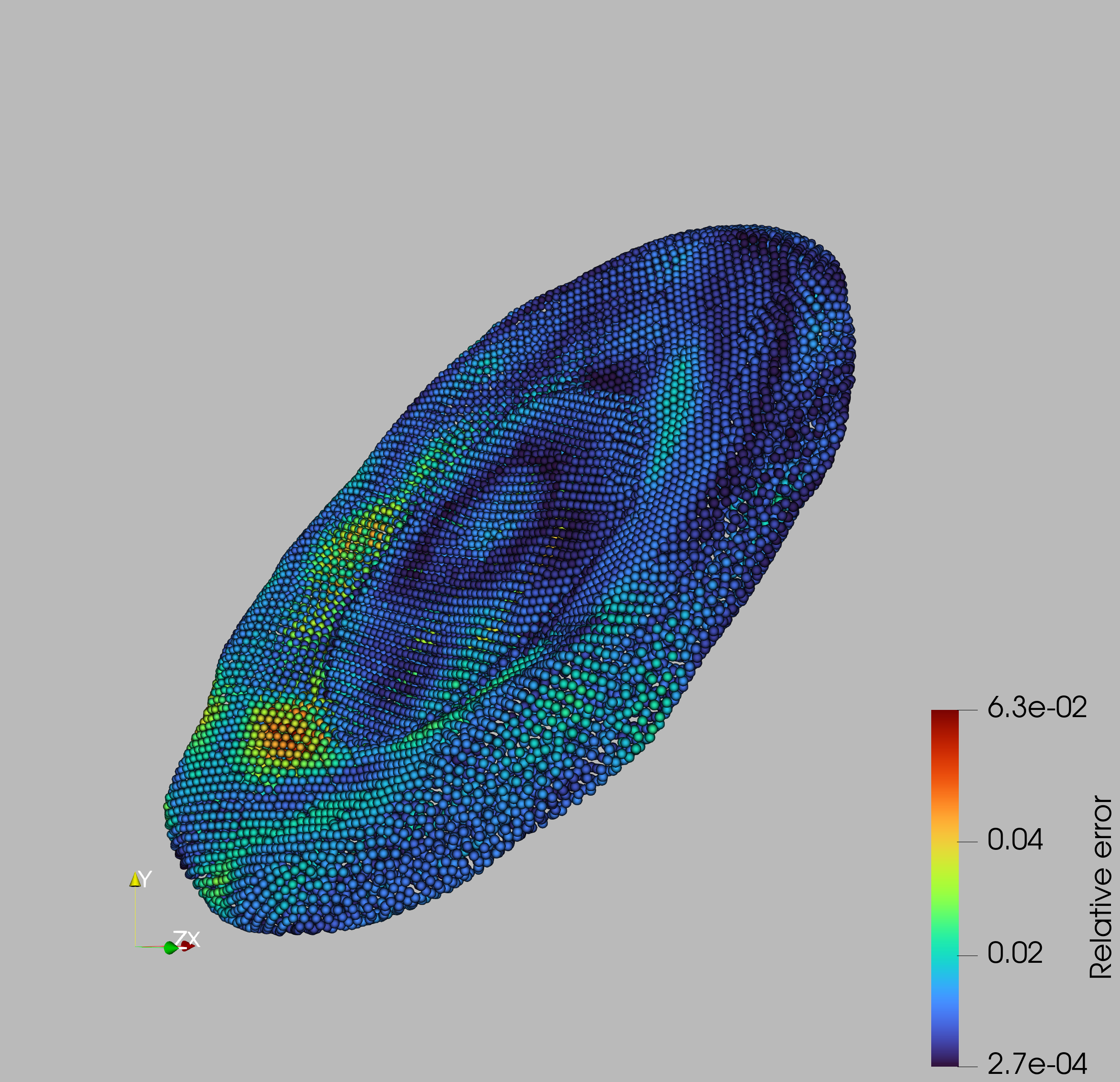}
    \caption{Per‐particle displacement comparison at the worst‐case capillary number: (a) ground‐truth magnitudes from LAMMPS, (b) DeepONet‐predicted magnitudes, and (c) corresponding point-wise relative error, illustrating the spatial distribution of surrogate discrepancies.}
    \label{fig:rel_err}
\end{figure}

A qualitative comparison of surrogate and ground‐truth displacement fields is shown in Fig.~\ref{fig:rel_err} for the worst‐case capillary number.
Panels (a) and (b) display the per–particle displacement magnitudes from the LAMMPS simulation and the DeepONet prediction, respectively, using the same color scale.
Panel (c) maps the point-wise relative error, revealing the spatial distribution of prediction discrepancies and highlighting the regions where the surrogate deviates most from the molecular‐dynamics reference.

\begin{figure}[h!]
    \centering
    \includegraphics[width=0.75\linewidth]{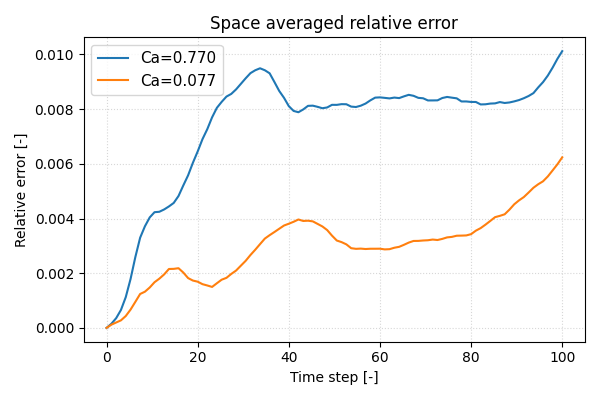}
    \caption{Space‐averaged relative displacement error of the DeepONet surrogate versus LAMMPS ground truth over one Jeffery period for two capillary numbers (\(Ca^{\ast}=0.77\) in blue, \(Ca^{\ast}=0.077\) in orange). Error peaks coincide with initial membrane self‐contact and rise again in the post‐contact, non‐periodic regime, highlighting regions of greatest model challenge.}
    \label{fig:error_in_time}
\end{figure}

To pinpoint when the surrogate deviates most from the molecular-dynamics reference, we plot the space-averaged relative displacement error versus time Fig.~\ref{fig:error_in_time} for the most complaint platelet ($Ca^*=0.77$) and the stiffest ($Ca^*=0.077$) one.
In both cases the error climbs rapidly when the inner faces of the hollow membrane first touch, a kinematic discontinuity that the network finds hardest to replicate.
Because the stiffer platelet deforms more slowly, this contact (and its associated error peak) occurs later in the trajectory.
Such self-contact is an artifact of the hollow-shell idealization; a realistic platelet, with cytoplasm and spectrin cytoskeleton, would not experience it, so the challenge to the surrogate should lessen in future models.
A secondary drift in error appears toward the end of the run, reflecting the non-periodic post-contact dynamics and the growing flow–structure complexity.
Incorporating physics-informed regularization, e.g. enforcing zero net hydrodynamic torque, offers a path to reduce this late-time error growth.

\begin{figure}[h!]
    \centering
    \includegraphics[width=0.75\linewidth]{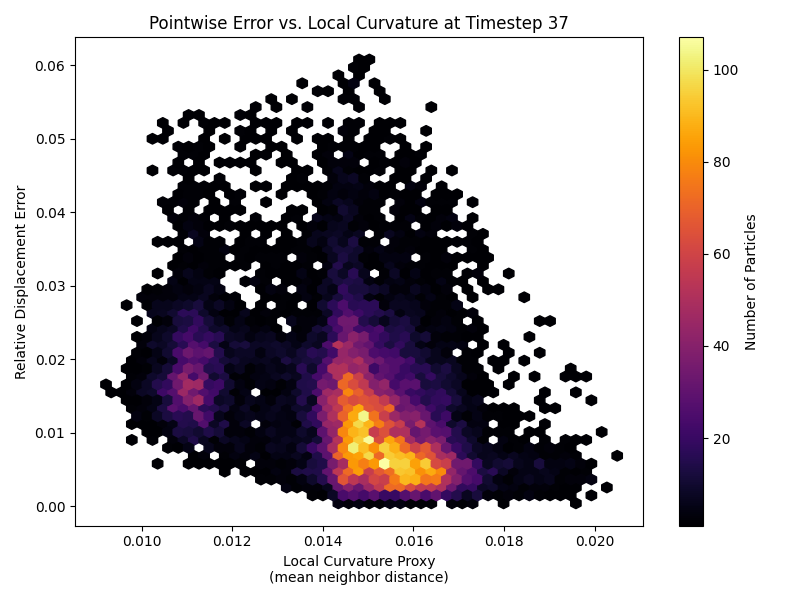}
    \caption{Hexagonal‐bin scatter of point-wise relative displacement error versus local curvature for the most compliant platelet (\(Ca^*=0.77\)) at first self‐contact (time step 37). Local curvature is proxied by the average distance to nearest neighbors, and bin color indicates the number of particles. The elevated error cluster at higher curvature values highlights the surrogate’s difficulty in capturing sharp bends and kinematic discontinuities.}
    \label{fig:curvature}
\end{figure}

At the moment of first self-contact, the spatial error distribution on the most compliant platelet ($Ca^*=0.77$) reveals a clear dependence on local membrane curvature (Fig.~\ref{fig:curvature}).
Each hexagon represents a patch of particles binned by its curvature proxy (mean distance to nearest neighbors) and point-wise relative displacement error.
A distinct cluster at higher curvature values exhibits elevated errors, indicating that regions with tight bends and kinematic discontinuities pose the greatest challenge for the fully connected DeepONet trunk net.
To better respect the platelet’s bead‐spring topology and improve accuracy in these high‐curvature zones, future work might replace the trunk net with a graph neural network that directly replicates the membrane graph structure.

\subsection{Extrapolation Study}
\label{ssec:extrapolation}
\begin{figure}[h!]
    \centering
    \includegraphics[width=0.75\linewidth]{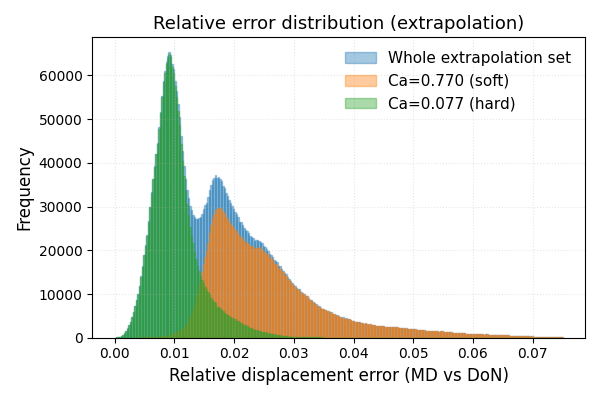}
    \caption{Relative displacement error histograms for the extrapolation study. The surrogate is evaluated on held‐out capillary numbers: the soft‐platelet case (\(Ca^*=0.77\), orange) and the stiff‐platelet case (\(Ca^*=0.08\), green), with the combined extrapolation set in blue.}
    \label{fig:extr_histo}
\end{figure}

The extrapolation study trains the DeepONet surrogate model by holding out the trajectory data for the two extreme $Ca^*$ cases,  
while keeping all training hyper-parameters identical to the baseline case.
Performance is then evaluated solely on the excluded capillary numbers.
Figure~\ref{fig:extr_histo} shows the resulting relative error histograms: a clear bimodal distribution arises, with the lower‐error peak corresponding to the stiff‐platelet extrapolation (green) and the higher‐error peak to the compliant‐platelet extrapolation (orange). In full extrapolation mode, most particles incur errors between 1\% and 3\%, and the maximum error remains below 8\%, underscoring the surrogate’s high-fidelity performance even outside its training envelope.

\section{Conclusions}
We have developed and bench-marked a DeepONet surrogate that reproduces the full, time–resolved deformation of a platelet in shear flow at a fraction of the cost of particle‐based simulation.
Training on ten capillary numbers spanning \(0.07\le Ca^{\ast}\le0.77\) and \(10^{6}\) particle states per trajectory, the network achieves sub-percent median errors and keeps the worst-case error below 4~\% across the entire calibrated range. 
A leave-extreme extrapolation test confirms that accuracy degrades gracefully: 90\% of the predictions for the held-out stiff and compliant platelets remain within 3\% of the LAMMPS reference, and the maximum error never exceeds 8\%.
These results place the surrogate firmly in the high-fidelity class while delivering speed-ups of four to five orders of magnitude—opening the door to organ-scale hemodynamic simulations with clinically realistic platelet counts.

Error analysis highlights two outstanding challenges.
First, localized peaks occur when the hollow membrane self-contacts, revealing the difficulty of capturing kinematic discontinuities with a fully connected trunk net.
Second, a slow drift appears in the non-periodic post-contact phase.
Both issues motivate future extensions: (i) replacing the trunk with a graph neural network that respects the bead–spring topology, and (ii) incorporating physics-informed regularization terms (such as zero net hydrodynamic torque) to constrain long-time behavior.
More realistic platelet geometries that include cytoplasm and cytoskeleton, thereby preventing self-contact, will further reduce surrogate error.

The present study establishes quantitative guidelines for deploying neural-operator surrogates inside multiscale thrombosis frameworks.
By combining continuum CFD with the demonstrated high-fidelity platelet model, forthcoming work will resolve clot initiation and growth in patient-specific geometries under physiologic shear, bridging molecular accuracy and vascular realism for the first time.


\begin{thebibliography}{10}

\bibitem{karniadakis2021physics}
George~Em Karniadakis, Ioannis~G Kevrekidis, Lu~Lu, Paris Perdikaris, Sifan Wang, and Liu Yang.
\newblock Physics-informed machine learning.
\newblock {\em Nature Reviews Physics}, 3(6):422--440, 2021.

\bibitem{lu2021learning}
Lu~Lu, Pengzhan Jin, Guofei Pang, Zhongqiang Zhang, and George~Em Karniadakis.
\newblock Learning nonlinear operators via deeponet based on the universal approximation theorem of operators.
\newblock {\em Nature machine intelligence}, 3(3):218--229, 2021.

\bibitem{laudato2024high}
Marco Laudato, Luca Manzari, and Khemraj Shukla.
\newblock High-fidelity description of platelet deformation using a neural operator.
\newblock {\em arXiv preprint arXiv:2412.00747}, 2024.

\bibitem{laudato2025neural}
Marco Laudato, Luca Manzari, and Khemraj Shukla.
\newblock Neural operator modeling of platelet geometry and stress in shear flow.
\newblock {\em arXiv preprint arXiv:2503.12074}, 2025.

\bibitem{macraild2024accelerated}
Michael MacRaild, Ali Sarrami-Foroushani, Toni Lassila, and Alejandro~F Frangi.
\newblock Accelerated simulation methodologies for computational vascular flow modelling.
\newblock {\em Journal of the Royal Society Interface}, 21(211):20230565, 2024.

\bibitem{xu2008multiscale}
Zhiliang Xu, Nan Chen, Malgorzata~M Kamocka, Elliot~D Rosen, and Mark Alber.
\newblock A multiscale model of thrombus development.
\newblock {\em Journal of the Royal Society Interface}, 5(24):705--722, 2008.

\bibitem{tomaiuolo2017regulation}
Maurizio Tomaiuolo, Lawrence~F Brass, and Timothy~J Stalker.
\newblock Regulation of platelet activation and coagulation and its role in vascular injury and arterial thrombosis.
\newblock {\em Interventional cardiology clinics}, 6(1):1, 2017.

\bibitem{zhang2002platelet}
Jian-ning Zhang, Angela~L Bergeron, Qinghua Yu, Carol Sun, Larry~V McIntire, Jos{\'e}~A L{\'o}pez, and Jing-fei Dong.
\newblock Platelet aggregation and activation under complex patterns of shear stress.
\newblock {\em Thrombosis and haemostasis}, 88(11):817--821, 2002.

\bibitem{zhang2016multi}
Yanhang Zhang, Victor~H Barocas, Scott~A Berceli, Colleen~E Clancy, David~M Eckmann, Marc Garbey, Ghassan~S Kassab, Donna~R Lochner, Andrew~D McCulloch, Roger Tran-Son-Tay, et~al.
\newblock Multi-scale modeling of the cardiovascular system: disease development, progression, and clinical intervention.
\newblock {\em Annals of biomedical engineering}, 44:2642--2660, 2016.

\bibitem{laudato2023buckling}
Marco Laudato, Roberto Mosca, and Mihai Mihaescu.
\newblock Buckling critical pressures in collapsible tubes relevant for biomedical flows.
\newblock {\em Scientific Reports}, 13(1):9298, 2023.

\bibitem{laudato2024analysis}
Marco Laudato and Mihai Mihaescu.
\newblock Analysis of the contact critical pressure of collapsible tubes for biomedical applications.
\newblock {\em Continuum Mechanics and Thermodynamics}, 36(1):217--228, 2024.

\bibitem{laudato2024sound}
Marco Laudato, Elias Zea, Elias Sundstr{\"o}m, Susann Boij, and Mihai Mihaescu.
\newblock Sound generation mechanisms in a collapsible tube.
\newblock {\em The Journal of the Acoustical Society of America}, 155(5):3345--3356, 2024.

\bibitem{gutierrez2024decoding}
Noelia~Grande Guti{\'e}rrez, Debanjan Mukherjee, and David Bark~Jr.
\newblock Decoding thrombosis through code: a review of computational models.
\newblock {\em Journal of Thrombosis and Haemostasis}, 22(1):35--47, 2024.

\bibitem{karmonik2008computational}
Christof Karmonik, Jean~X Bismuth, Mark~G Davies, and Alan~B Lumsden.
\newblock Computational hemodynamics in the human aorta: a computational fluid dynamics study of three cases with patient-specific geometries and inflow rates.
\newblock {\em Technology and Health Care}, 16(5):343--354, 2008.

\bibitem{sundstrom2023machine}
Elias Sundstr{\"o}m and Marco Laudato.
\newblock Machine learning-based segmentation of the thoracic aorta with congenital valve disease using mri.
\newblock {\em Bioengineering}, 10(10):1216, 2023.

\bibitem{bornemann2024relation}
Karoline-Marie Bornemann, Silje~Ekroll Jahren, and Dominik Obrist.
\newblock The relation between aortic morphology and transcatheter aortic heart valve thrombosis: Particle tracing and platelet activation in larger aortic roots with and without neo-sinus.
\newblock {\em Computers in biology and medicine}, 179:108828, 2024.

\bibitem{zhang2021predictive}
Peng Zhang, Jawaad Sheriff, Shmuel Einav, Marvin~J Slepian, Yuefan Deng, and Danny Bluestein.
\newblock A predictive multiscale model for simulating flow-induced platelet activation: Correlating in silico results with in vitro results.
\newblock {\em Journal of biomechanics}, 117:110275, 2021.

\bibitem{wang2023multiscale}
Peineng Wang, Jawaad Sheriff, Peng Zhang, Yuefan Deng, and Danny Bluestein.
\newblock A multiscale model for shear-mediated platelet adhesion dynamics: correlating in silico with in vitro results.
\newblock {\em Annals of Biomedical Engineering}, 51(5):1094--1105, 2023.

\bibitem{gupta2021multiscale}
Prachi Gupta, Peng Zhang, Jawaad Sheriff, Danny Bluestein, and Yuefan Deng.
\newblock A multiscale model for multiple platelet aggregation in shear flow.
\newblock {\em Biomechanics and modeling in mechanobiology}, 20:1013--1030, 2021.

\bibitem{kruger2016effect}
Timm Kr{\"u}ger.
\newblock Effect of tube diameter and capillary number on platelet margination and near-wall dynamics.
\newblock {\em Rheologica Acta}, 55:511--526, 2016.

\bibitem{liao2022flow}
Chih-Tang Liao, An-Jun Liu, and Yeng-Long Chen.
\newblock Flow-induced “waltzing” red blood cells: microstructural reorganization and the corresponding rheological response.
\newblock {\em Science Advances}, 8(47):eabq5248, 2022.

\bibitem{abidin2023microfluidic}
Nurul A~Zainal Abidin, Mariia Timofeeva, Crispin Szydzik, Farzan Akbaridoust, Chitrarth Lav, Ivan Marusic, Arnan Mitchell, Justin~R Hamilton, Andrew~SH Ooi, and Warwick~S Nesbitt.
\newblock A microfluidic method to investigate platelet mechanotransduction under extensional strain.
\newblock {\em Research and Practice in Thrombosis and Haemostasis}, 7(1):100037, 2023.

\bibitem{malipeddi2021shear}
Abhilash~Reddy Malipeddi and Kausik Sarkar.
\newblock Shear-induced gradient diffusivity of a red blood cell suspension: effects of cell dynamics from tumbling to tank-treading.
\newblock {\em Soft Matter}, 17(37):8523--8535, 2021.

\bibitem{vahidkhah2013hydrodynamic}
Koohyar Vahidkhah, Scott~L Diamond, and Prosenjit Bagchi.
\newblock Hydrodynamic interaction between a platelet and an erythrocyte: effect of erythrocyte deformability, dynamics, and wall proximity.
\newblock {\em Journal of biomechanical engineering}, 135(5):051002, 2013.

\bibitem{dynar2024platelet}
Mariam Dynar, Hamid Ez-Zahraouy, Chaouqi Misbah, and Mehdi Abbasi.
\newblock Platelet margination dynamics in blood flow: The role of lift forces and red blood cells aggregation.
\newblock {\em Physical Review Fluids}, 9(8):083603, 2024.

\bibitem{tuna2024platelet}
Rukiye Tuna, Wenjuan Yi, Esmeralda Crespo~Cruz, JP~Romero, Yi~Ren, Jingjiao Guan, Yan Li, Yuefan Deng, Danny Bluestein, Zixiang~Leonardo Liu, et~al.
\newblock Platelet biorheology and mechanobiology in thrombosis and hemostasis: Perspectives from multiscale computation.
\newblock {\em International Journal of Molecular Sciences}, 25(9):4800, 2024.

\bibitem{thompson2022lammps}
Aidan~P Thompson, H~Metin Aktulga, Richard Berger, Dan~S Bolintineanu, W~Michael Brown, Paul~S Crozier, Pieter~J In't~Veld, Axel Kohlmeyer, Stan~G Moore, Trung~Dac Nguyen, et~al.
\newblock Lammps-a flexible simulation tool for particle-based materials modeling at the atomic, meso, and continuum scales.
\newblock {\em Computer physics communications}, 271:108171, 2022.

\bibitem{groot1997dissipative}
Robert~D Groot and Patrick~B Warren.
\newblock Dissipative particle dynamics: Bridging the gap between atomistic and mesoscopic simulation.
\newblock {\em The Journal of chemical physics}, 107(11):4423--4435, 1997.

\bibitem{zhang2014multiscale}
Peng Zhang, Chao Gao, Na~Zhang, Marvin~J Slepian, Yuefan Deng, and Danny Bluestein.
\newblock Multiscale particle-based modeling of flowing platelets in blood plasma using dissipative particle dynamics and coarse grained molecular dynamics.
\newblock {\em Cellular and molecular bioengineering}, 7:552--574, 2014.

\end{thebibliography}

\end{document}